\documentclass[twocolumn,english,prd]{revtex4}
\usepackage{graphicx}
\newcommand{\eeq}{\end{equation}}
\newcommand{\beqa}{\begin{eqnarray}}
\newcommand{\eeqa}{\end{eqnarray}}
\newcommand{\fnl}{f_{\rm NL}}
\newcommand{\Phip}{\Phi^{\rm prim}}
\def\Mpc{\, h^{-1} \, {\rm Mpc}}
\def\Gpc{\, h^{-1} \, {\rm Gpc}}
\def\kMpc{\, h \, {\rm Mpc}^{-1}}

\def\fun#1#2{\lower3.6pt\vbox{\baselineskip0pt\lineskip.9pt
        \ialign{$\mathsurround=0pt#1\hfill##\hfil$\crcr#2\crcr\sim\crcr}}}

\def\la{\mathrel{\mathpalette\fun <}}
\def\ga{\mathrel{\mathpalette\fun >}}

\newcommand{\mnras}{Mon. Not. R. Astron. Soc.}
\newcommand{\aap}{Astron. Astrophys.}

\newcommand{\beq}{\begin{equation}}
\newcommand{\enq}{\end{equation}}
\newcommand{\bdis}{\begin{displaymath}}
\newcommand{\edis}{\end{displaymath}}
\newcommand{\kvec}{{\bf k}}
\newcommand{\q}{{\bf q}}
\newcommand{\de}{{\rm d}}

\begin{document}

\title{Probing Primordial Non-Gaussianity with Large-Scale Structure}

\author{Rom\'an Scoccimarro$^1$, Emiliano Sefusatti$^1$, Matias Zaldarriaga$^2$}
 \vskip 2pc
\address{${}^{1}$Center for Cosmology and Particle Physics\\
Department of Physics, New York University \\
New York, NY 10003\\
${}^{2}$Department of Astronomy and Department of Physics\\
Harvard University, 60 Garden Street\\
Cambridge, MA 02138}

\begin{abstract}

We consider primordial non-Gaussianity due to quadratic corrections in the gravitational potential parametrized by a non-linear coupling parameter $\fnl$. We study constraints on $\fnl$ from measurements of the galaxy bispectrum in redshift surveys. Using estimates for idealized survey geometries of the 2dF and SDSS surveys and realistic ones from SDSS  mock catalogs, we show that it is possible to probe $|\fnl| \simeq 100$, after marginalization over bias parameters. We apply our methods to the  galaxy bispectrum measured from the PSCz survey, and obtain a $2\sigma$-constraint $|\fnl|\la 1800$. We estimate that an all sky redshift survey up to $z\simeq 1$ can probe $|\fnl| \simeq 1$. We also consider the use of cluster abundance to constrain $\fnl$ and find that in order to be sensitive to $|\fnl| \simeq 100$, cluster masses need to be determined with an accuracy of a few percent, assuming perfect knowledge of the mass function and cosmological parameters.
\end{abstract}

\maketitle

\section{Introduction}

Rapid progress in microwave background anisotropy experiments and large galaxy redshift surveys is providing high quality data that can be used to test the nature of primordial fluctuations. The
leading scenario for explaining the initial seeds for the
formation of structure in the universe is inflation, a period of
accelerated expansion in the early universe during which quantum
fluctuations in a scalar field driving the expansion  are
stretched outside the Hubble radius and stay frozen until they
cross back during matter domination and grow by gravitational
instability. The predictions from this scenario have been worked
out in great detail during the last twenty years, with most models of
inflation predicting a scale-invariant spectrum of adiabatic
Gaussian fluctuations. 

The Gaussian nature of primordial perturbations is a direct consequence of
the slow-roll conditions on the inflation potential, required for
the potential energy to dominate over the kinetic energy of the field and produce
a sufficiently long period of  accelerated expansion. Under these circumstances,  
non-Gaussianities are very small, of the order of the tilt in the scalar
spectrum~\cite{Malda03}. The tilt  is known
to be rather small~\cite{Spergel:2003cb,Peiris:2003ff}, in the
language of Eq.~(\ref{fnleq}) below,  $\fnl \la 0.05$. This bound can
be relaxed to $\fnl \sim 1$ if 
higher dimensional operators are suppressed by the lowest 
possible scale consistent with slow-roll~\cite{paolo}.

There are several modifications to the basic 
physics of inflation that can lead to larger non-Gaussianities. 
In some set of these models there are additional light degrees of freedom 
during inflation.  
For example, if the inflaton field has more than one component, it is
possible to generate significant  non-Gaussianity in the adiabatic
component through the coupling to isocurvature components, see
e.g.~\cite{BeUz02,BMR02,BeUz03}. A second scalar field, 
usually called the ``curvaton'' could be light during inflation 
and come to dominate  the energy density of 
the universe after the end of inflation before producing effectively 
a second reheating~\cite{mollerach,linde2,lyth,moroi,enqvist,bartolo}. 
The density fluctuations we observe today could be due to fluctuations in 
the curvaton and could be non-Gaussian. 

In addition, recent
work~\cite{DGZ03, Zalda03,DGZ03b}  suggested a new 
possibility in which the fluctuations are generated
during the reheating period when the inflaton energy density is
converted into standard model particles with a fluctuating decay
rate.  In all these models  non-Gaussianities are primarily as 
given by  Eq.~(\ref{fnleq}) and could naturally be of order 
 $\fnl \approx 5-30$~\cite{Zalda03,DGZ03b}. Primordial non-Gaussianity at this
level should be detectable through measurements of the bispectrum
of the CMB~\cite{Komatsu:rj,Komatsu:2003fd} and in galaxy surveys,
as we shall show in this paper. 
Finally there are models in which the 
inflaton is not a slowly rolling scalar field but rather a fast moving ghost condensate~\cite{nima}. In this case non-Gaussianities are much larger, close to the current 
upper limit  but are not as simple as Eq.~(\ref{fnleq}). 

In this paper we consider departures of
Gaussianity where the primordial gravitational potential at
sub-horizon scales has the form~\cite{SaBo} \beq \label{fnleq}
\Phip = \phi+\frac{\fnl}{c^2}(\phi^2- \langle\phi^2\rangle) \enq
where $\phi$ is a random Gaussian field, $c$ is the speed of
light, and for simplicity we assume $\fnl$ is a number independent
of scale. These are generically predicted by all models in which the non-Gaussianities are generated outside the horizon~\cite{Zalda03}.  

Measurements of the microwave background anisotropy
bispectrum give $1\sigma$ limits from COBE $|\fnl| \leq
1500$~\cite{COBEfnl}, MAXIMA $|\fnl| \leq 950$~\cite{MAXIMAfnl},
and recently $2\sigma$ limits from WMAP of $-58\leq \fnl \leq
134$~\cite{Komatsu:2003fd}. Upon completion, WMAP is expected to
reach $1\sigma$ sensitivity of order $|\fnl|\la20$, whereas the
Planck satellite would yield $|\fnl|\la5$~\cite{Komatsu:rj}.

An alternative way of constraining primordial non-Gaussianity is by measuring the bispectrum of the galaxy distribution and looking for deviations from the predictions of gravitational instability from Gaussian initial conditions~\cite{FrSc94,Sco00, VWHK00} (see e.g.~\cite{PTReview} for a review). In this case, there has been no constraint yet on the particular model given by Eq.~(\ref{fnleq}), although estimates have been made in~\cite{VWHK00} regarding the ability of large-scale structure to constrain $\fnl$, concluding that galaxy surveys such as 2dF and SDSS will be able to probe only $|\fnl| \sim 10^3-10^4$, and that galaxy surveys in general will not be competitive with CMB experiments in probing this type of non-Gaussianity.

In this paper, we revisit the issue of how well large-scale structure can constraint non-Gaussianity of the type given by Eq.~(\ref{fnleq}), and reach a quite different conclusion. We show in particular that the SDSS galaxy bispectrum should be able to probe values of order $|\fnl| \sim 10^2$, and that a hypothetical all-sky survey with similar density up to $z \sim 1$ should be able to probe $|\fnl| \sim 1$.  We illustrate our results by applying these ideas to the measurement of the galaxy bispectrum in the PSCz survey~\cite{FFFS01}, and find $2\sigma$ constraints $|\fnl| \la 1800$, comparable to the limits from CMB measurements before WMAP. 

The difference between our results and those in~\cite{VWHK00} can be traced to the assumption made in~\cite{VWHK00} that constraints on primordial non-Gaussianity can be effectively ``read off"  from constraints on the non-linear bias parameter, which is independent of scale. This ignores the anomalous scale dependence of the bispectrum induced by primordial non-Gaussianity, which plays a crucial role in obtaining limits on it~\cite{SFFF01, FFFS01}.

Constraints on primordial non-Gaussianity of the type given by Eq.~(\ref{fnleq}) have been considered also by using gravitational lensing~\cite{TaJa03}, where it was found that it is possible to achieve $\fnl \approx 150 f_{\rm sky}^{-1/2}$, with $f_{\rm sky}$ the fraction of sky covered, using lensing tomography with 4 redshift bins up to $\ell=500$. In addition, the use of the abundance of massive clusters to constrain $\fnl$ has been studied in~\cite{Matarrese:2000iz,Verde:2000vr}. In this work we also consider how well it is necessary to determine cluster masses to be able to use cluster abundance to probe $\fnl$ to the accuracy required by present upper limits. 

This paper is organized as follows. In section~\ref{bisplss} we discuss how the large-scale structure (LSS) bispectrum is modified due to primordial non-Gaussianity given by Eq.~(\ref{fnleq}). Section~\ref{bispStoN} presents a signal-to-noise analysis for determining bias parameters and $\fnl$ from surveys, including a somewhat detailed calculation for the particular case of the SDSS survey, and application to the bispectrum of galaxies in the PSCz survey (Section~\ref{pscz}). Finally, in  Section~\ref{clus} we consider the use of the abundance of clusters to constrain primordial non-Gaussianity.

\section{The LSS Bispectrum}
\label{bisplss}

Well after the universe becomes matter dominated, the fluctuations in the gravitational potential at time given by the scale factor $a$ are related to the primordial fluctuations by

\beq
\Phi_{\kvec}(a)=\frac{9}{10}\frac{D_+}{a}T(k)\ \Phip_{\kvec},
\enq
where $T(k)$ is the transfer function, $D_+(a)$ is the growth factor linear perturbation theory, and $\Phip$ denotes the primordial gravitational potential at sub-horizon scales before matter-radiation transition, and we have neglected anisotropic stresses. The matter density is related to the potential by Poisson's equation, which in Fourier space reads
\beq
\delta_{\kvec}(a)=-\frac{2}{3}\frac{a\ k^2}{\Omega_mH_0^2}\Phi_{\kvec}(a),
\enq
where $\Omega_m$ is the present value of the dark matter density in terms of the critical density and $H_0=100 \kMpc {\rm km/s}$ is the present value of the Hubble constant. We assume cosmological parameters consistent with those determined by the WMAP experiment~\cite{Spergel:2003cb}, $\Omega_m=0.27$, $\Omega_b h^2=0.0224$, $h=0.71$, assuming a flat universe with a cosmological constant. The transfer function is computed using CMBFAST~\cite{SeZa96}, leading to a power spectrum normalization $\sigma_8=0.82$. Introducing
\beq
M(k,a)\equiv -\frac{3}{5}\frac{k^2T(k)}{\Omega_mH_0^2}D_+(a)
\enq
we can write
\beq
\delta_{\kvec}(a)=M(k,a)\ \Phip_{\kvec},
\label{dephi}
\enq
and in general
\begin{eqnarray}\label{npoint}
\langle\delta_{\kvec_1}\delta_{\kvec_2}... \delta_{\kvec_N}\rangle=M(k_1)M(k_2)... M(k_N)\times\nonumber\\
\langle\Phip_{\kvec_1}\Phip_{\kvec_2}... \Phip_{\kvec_N}\rangle
\end{eqnarray}
Henceforth we shall suppress the dependence on the scale factor, assuming $a=1$, and drop the superscript denoting the primordial gravitational potential, which is understood in all our expressions that follow. In our convention the power spectrum and the bispectrum are given by
\begin{eqnarray}
\langle\delta_{\kvec_1}\delta_{\kvec_2}\rangle & \equiv & \delta_D (\kvec_{12})\  P(k_1)\\
\langle\delta_{\kvec_1}\delta_{\kvec_2}\delta_{\kvec_3}\rangle & \equiv & \delta_D (\kvec_{123})\ B(k_1,k_2,k_3)
\end{eqnarray}
where $\kvec_{i\ldots j}\equiv\kvec_i+\ldots+\kvec_j$.  We can write the linear power spectrum of the density field as $P^L(k)=M^2(k) P_\Phi(k)$ where $P_\Phi(k)$ is the \textit{primordial} gravitational potential power spectrum, given by Eq.~(\ref{fnleq})
\begin{eqnarray}
P_\Phi(k) & = & P_\phi(k)+2 \frac{\fnl^2}{c^4}\int d^3q P_\phi(q)P_\phi(|\kvec-\q|)\nonumber\\
& \simeq & P_\phi(k)
\label{Pphi}
\end{eqnarray}
Equation~(\ref{fnleq}) can be seen as a quadratic approximation to a more general power series expansion, for consistency we neglect higher-order corrections than those of leading order in the primordial non-Gaussianity parameter $\fnl$. For example, the second-order correction in Eq.~(\ref{Pphi}) changes the value of $\sigma_8$ by less than 1$\%$ for $\fnl=10^2$.
From Eq.~(\ref{npoint}), it  follows that the bispectrum in linear perturbation theory is given by
\beq\label{initialbispectrum}
B^L_{123}=M(k_1)M(k_2)M(k_3)\ B_\Phi(k_1,k_2,k_3),
\enq
where $B_{123}\equiv B(k_1,k_2,k_3)$ and
\begin{eqnarray}
B_\Phi(k_1,k_2,k_3) & = & \frac{2 \fnl}{c^2} \left[P_\phi(k_1)P_\phi(k_2)+\rm{cyc.}\right]\nonumber\\
& & +{\mathcal O}(\fnl^3).
\label{bphi}
\end{eqnarray}

\begin{figure}
\begin{center}
\includegraphics[width=0.5\textwidth]{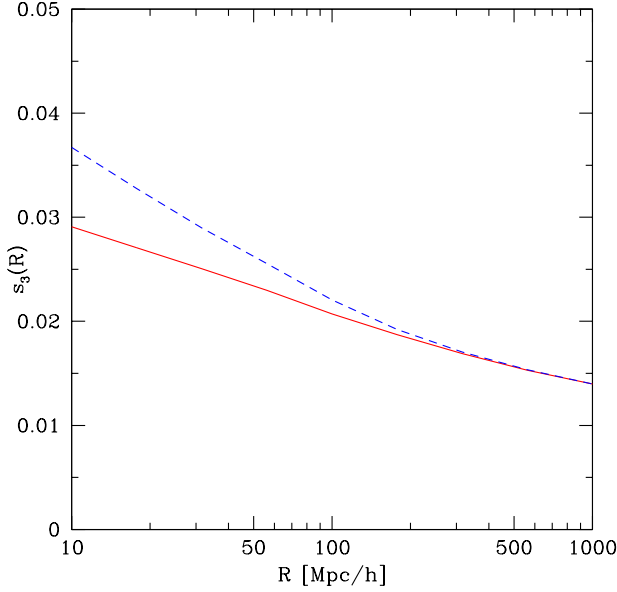}
\caption{\label{skew}The dimensionless skewness parameter $s_3(R)$ against smoothing scale $R$ for $\fnl=-100$ (solid line) and the approximation given by Eq.~(\ref{s3app}) (dashed line).}
\end{center}
\end{figure}

To get a sense of how significant is primordial non-Gaussianity for the {\em density field} as a function of $\fnl$, we calculate the dimensionless skewness parameter, defined as
\beq
s_3(R)\equiv \frac{\langle\delta_R^3\rangle}{\langle\delta_R^2\rangle^{3/2}},
\label{s3png}
\enq
where the smoothed density field $\delta_R$ is given by
\beq
\delta_R = \int d^3k\ W(kR)\ \delta_\kvec,
\label{deltaR}
\eeq
where $W(kR)$ denotes the Fourier transform of a spherical top-hat window of radius $R$ in real space. From Eqs.~(\ref{dephi},\ref{bphi}-\ref{s3png}) we have
\beqa
s_3(R)&=&\frac{6\fnl}{c^2}\langle \delta_R^2\rangle^{-3/2} \int d^3k_1\, d^3k_2\, P_\phi(k_1) P_\phi(k_2) \nonumber \\
& & \times M_1M_2M_{12}\ W_1W_2W_{12},
\label{s3R}
\eeqa
where $M_i\equiv M(k_i)$ and $W_i\equiv W(k_i R)$ and $k_{12}=|\kvec_1+\kvec_2|$. We can integrate this equation numerically~\cite{Matarrese:2000iz}, but it is also possible to derive an analytic expression that is exact at large scales and illustrates the basic dependence of $s_3(R)$ on cosmological parameters. The non-trivial part in Eq.~(\ref{s3R}) is the integration over the angle between $\kvec_1$ and $\kvec_2$, due to the dependence of $M_{12}$ on the transfer function $T(k_{12})$. We use the approximation,

\beqa
\int \frac{d\Omega_{12}}{4\pi}\ W_{12}\, k_{12}^2\, T_{12}& \approx &k_1^2 T_1 W_1 \left( W_2+\frac{k_2R}{3}W_2'\right) \nonumber \\ & & + 1 \leftrightarrow 2,
\label{app}
\eeqa
which is exact at large scales where $T(k)$ is independent of $k$, in view of the summation theorem of Bessel functions (see Appendix C in~\cite{PTReview}). After simple algebra, Eq.~(\ref{app}) leads to   
\beq
s_3(R)\approx 12\, \frac{f_{\rm NL}}{c^2}\, \frac{\langle \delta_R
\phi_R\rangle}{\sqrt{\langle \delta_R^2 \rangle}} \left[1+\frac{1}{6}
\frac{d\ln\langle \delta_R\phi_R\rangle}{d\ln R}\right].
\label{s3app}
\eeq
This formula illustrates that the level of non-Gaussianity is proportional to $\fnl$ times the amplitude of potential fluctuations smoothed on scale $R$, with a constant of proportionality that depends on the shape of the density-potential power spectrum.  Figure~\ref{skew}  shows a plot of $s_3(R)$ as a function of scale $R$ calculated by numerically integrating Eq.~(\ref{s3R}) (solid line) and the analytic expression in Eq.~(\ref{s3app}), which matches the exact result at large scales.

At the scales relevant for galaxy surveys, we must take into account second-order corrections in perturbation theory (PT), which read
\beq
\delta_{\kvec} \simeq \delta_{\kvec}^L +\int d^3q\ F_2(\q,\kvec-\q)\ \delta_{\q}^L \delta_{\kvec-\q}^L
\enq
where ($x \equiv \hat{k}_1 \cdot \hat{k}_2$)
\beq
F_2(\kvec_1,\kvec_2) = \frac{5}{7}+\frac{x}{2}\left(\frac{k_1}{k_2}+\frac{k_2}{k_1}\right)
+\frac{2}{7}\ x^2.
\label{F2}
\eeq
This gives for the power spectrum,
\beqa
P(k)&=&P^L(k)+2\int d^3q\, F_2(\kvec+\q,-\q)\, B^L(\kvec,\q) \nonumber \\
& \equiv & P^L(k)+P^B(k),
\label{Pkng}
\eeqa
however, this correction is basically negligible at the scales we are interested, see Eq.~(\ref{PNGPL}) below. For the bispectrum we have three contributions~\cite{FrSc94},
\beq
B_{123}=B^L_{123}+B^G_{123}+B^T_{123}
\enq
of which $B^L$ is the linearly evolved bispectrum due to primordial fluctuations, $B^T$ is the contribution due to the initial trispectrum, and $B^G$ is the usual bispectrum generated by gravitational instability from Gaussian initial conditions. They are given by
\begin{eqnarray}
B^G_{123} & = & 2 F_2(\kvec_1,\kvec_2)P_1 P_2+\;\text{cyc.}\\ & & \nonumber
\label{BG} \\
B^T_{123} & = & \int d^3 q F_2(\kvec_{12}-\q,\q)T^L(\kvec_1,\kvec_2,\kvec_{12}-\q,\q)\nonumber\\
& & +\;\rm{cyc.} \label{BT}
\end{eqnarray}
where $P_i\equiv P(\kvec_i)$. The linearly evolved initial trispectrum $T^L$ is a quantity of second order in $\fnl$ and can therefore be ignored as we now show. We can estimate the corrections due to primordial non-Gaussianity for an equilateral triangle of side $k$ (at large enough scales),
\beqa
B^L &= & \frac{7}{2} \varepsilon \ B^G
\label{BLBG}\\
B^T &\simeq &- 3 \varepsilon^2 \ k^2\sigma_v^2  \ B^G
\label{BTBL}\\
P^B &\simeq &-\frac{4}{7} \varepsilon\ k^2\sigma_v^2  \ P^L
\label{PNGPL}
\eeqa

where $\sigma_v^2$ is the one-dimensional velocity dispersion in units of the Hubble constant,

\beq
\sigma_v^2 = \frac{1}{3} \int d^3q\ \frac{P(q)}{q^2} \simeq  40\ (\Mpc)^2,
\eeq
and the small parameter $\varepsilon$ is given by

\beq
\varepsilon = \frac{\fnl}{M(k)c^2} \simeq -0.07\, \frac{\fnl}{100}\, \frac{1}{T(k)} \left(\frac{0.01 \kMpc}{k}\right)^2,
\label{eps}
\eeq
which leads to
\beq
\varepsilon\, k^2 \sigma_v^2 \simeq 3\times 10^{-4}\ \frac{\fnl}{100}\ \frac{1}{T(k)}.
\eeq

Equation~(\ref{PNGPL}) says that the power spectrum correction due to primordial non-Gaussianity, Eq.~(\ref{Pkng}), is suppressed by $\varepsilon\, k^2 \sigma_v^2$ at large scales, and becomes of order a few percent as $k \ga 0.1 \kMpc$, inducing a scale dependence on the bispectrum, see Fig.~\ref{reducedB} below. At these scales other effects due to non-linearities (specially redshift distortions) become important, we found that in redshift space it is very difficult to see the effects of nonzero $\fnl$ at these scales. Equation~(\ref{BTBL}) shows the trispectrum correction given by Eq.~(\ref{BT}) is negligible. Therefore we only probe primordial non-Gaussianity through Eq.~(\ref{BLBG}).

It is convenient to introduce the \textit{reduced bispectrum}~\cite{Fry84b}, defined as
\beq\label{defQ}
Q_{123}\equiv\frac{B_{123}}{P_1 P_2+P_1 P_3+P_2 P_3},
\enq
which is independent of time for Gaussian initial conditions; moreover, in tree-level perturbation theory it reduces to a scale independent value for equilateral configurations~\cite{Fry84b}

\beq
Q_{\rm eq}^G(k)= \frac{B^G(k,k,k)}{3\ [P(k)]^2}=\frac{4}{7}.
\label{QeqG}
\enq

For general triangles, $Q$ retains approximately this simple behavior, it is independent of power spectrum normalization, and only very weakly dependent on $\Omega_m$ through the factor $\simeq \Omega_m^{-2/63}$~\cite{BCHJ95}, the only relevant dependence of $Q$ is on the local spectral index $n_{\rm eff}(k) \equiv \de\ln P/\de\ln k$ and triangle shape through Eq.~(\ref{F2}).

An important consequence of primordial non-Gaussianity from Eq.~(\ref{fnleq}) is that it violates the scaling induced by gravity, since $Q_{\rm eq}^L(k) \sim \varepsilon \sim 1/[k^2T(k)]$, see Eq.~(\ref{eps}). The top panel in Fig.~\ref{reducedB} illustrates the deviations from Eq.~(\ref{QeqG}) when $|\fnl|=10^2$. Note that the scale dependence seen here is opposite to that in the skewness (compare to Fig.~\ref{skew}); this is simply due to the difference in normalizations between Eq.~(\ref{s3png}) and~(\ref{defQ}). The bottom panel in Fig.~\ref{reducedB} shows the corrections due to primordial non-Gaussianity as a function of triangle shape for $k_1=0.02\kMpc$ and $k_2=2 k_1$. We now explore how well these deviations can be probed with galaxy surveys.

\begin{figure}
\begin{center}
\includegraphics[width=0.5\textwidth]{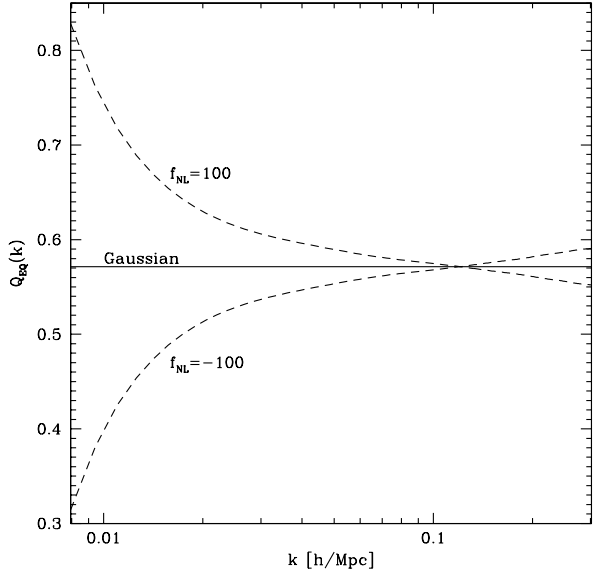}
\includegraphics[width=0.5\textwidth]{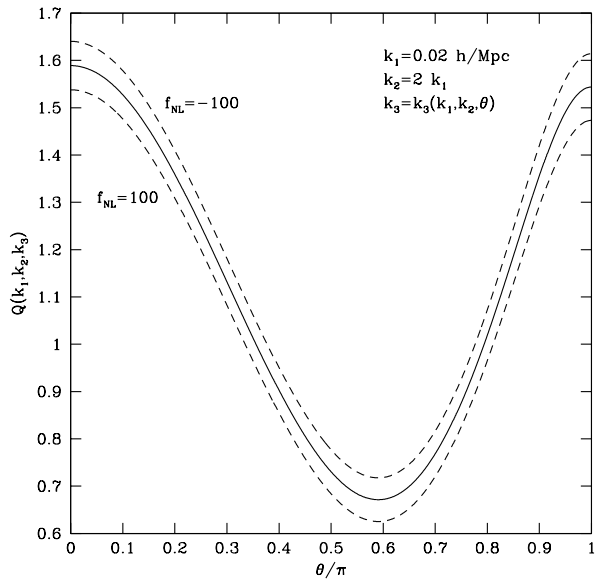}
\caption{Primordial Non-Gaussianity corrections to the reduced bispectrum for equilateral configurations as a function of the wavenumber $k$ (top panel) and for $k_1=0.02\kMpc$ and $k_2=2 k_1$ as a function of the angle $\theta$ between $\kvec_1$ and $\kvec_2$ (bottom).}\label{reducedB}
\end{center}
\end{figure}

\section{Bispectrum Signal to Noise}
\label{bispStoN}

\subsection{Order of Magnitude Estimate}

In this section we will calculate the minimum $\fnl$ that can be measured  by a  galaxy redshift survey as a function of the survey parameters. We start by making a simple estimate to put the results of the next sections in context. A survey with volume $V$  contains 
\beqa
N_k&\sim& {4\pi \over 3}\, k_{\max}^3\  \frac{V}{(2\pi)^3} \nonumber \\
&\sim& 4.5 \times 10^5 \ \frac{V}{(1 \Gpc)^3}\ \left(\frac{k_{\max}}{0.3 \kMpc}\right)^3 \nonumber \\ & &
\eeqa
Fourier modes, where  $k_{\max}$ is the largest wavenumber that can be used in the analysis. 
Let us use the skewness to estimate $\fnl$. With this number of modes we expect to be able to measure the skewness roughly to the level, 
\beq
\Delta s_3 \sim \sqrt{15 \over N_k},
\eeq
where we have used the Gaussian variance. In the previous section we showed that the skewness is of order $s_3 \sim 2 \times 10^{-4} \fnl$ so we expect to be able to detect $\fnl\sim 30$ if $V\sim  (1 \ h^{-1} {\rm Gpc})^3$. 

The above estimate indicates that we expect LSS surveys to be competitive with CMB experiments in  constraining primordial non-Gaussianity. However, there are several simplifications in this estimate: 1) The density field fluctuations are not Poisson distributed, thus the skewness is not the best estimate of $\fnl$; 2) The galaxy density field is a biased tracer of the underlying mass and biasing modifies the bispectrum, therefore one must determine simultaneously $\fnl$ and bias parameters; 3) The survey geometry can significantly complicate the determination of the bispectrum. We will tackle these problems in the rest of this section to obtain a more robust estimate of the capabilities of redshift surveys. 

\subsection{Ideal Geometry}

We first consider the case of ideal survey geometry, assuming that bispectra for different triangle shapes are uncorrelated, i.e. the bispectrum covariance matrix is diagonal and given by Gaussian statistics.
In order to see how well one can probe non-Gaussianity one has to include also the possibility that galaxies are biased tracers of the density field. At large scales, it is reasonable to assume that biasing is local, then~\cite{FrGa93}

\beq
\delta_g = b_1\ \delta + \frac{b_2}{2}\ \delta^2+ \ldots
\enq
where $b_1$ and $b_2$ are constants. The bispectrum in the galaxy distribution, including primordial  non-Gaussianity, will be given by
\begin{eqnarray}\label{bias_bisp}
B_g(k_1,k_2,k_3) & = & b_1^3\ B^G_{123}+b_1^2 b_2\ (P_1 P_2+ \textrm{cyc.})+\nonumber\\
& & +b_1^3\ B^L_{123}
\end{eqnarray}
In terms of the galaxy reduced bispectrum $Q_g$, we have
\beq
Q_g=\frac{Q^G_{123}}{b_1}+\frac{b_2}{b_1^2}+ \frac{Q^L_{123}}{b_1}.
\label{Qgtot}
\enq
Note that each term in this expression has a different behavior, $Q^G_{123}$ depends very weakly on scale through the local spectral index (which can be measured) and depends strongly on triangle configuration, the second term due to non-linear bias is a constant, and the last term due to primordial non-Gaussianity depends rather strongly on scale (see Fig.~\ref{reducedB}). Therefore it is possible to simultaneously obtain constraints on $b_1$, $b_2$ and $\fnl$.

For the reasons discussed above, we work with the reduced bispectrum $Q_{123}$~\footnote{From now on we refer to $Q$ as the bispectrum, rather than reduced bispectrum.}, which has identical  signal to noise to $B_{123}$ in the limit of Gaussian fluctuations. Indeed, for $Q_{123}=B_{123}/\Sigma_{123}$, where $\Sigma_{123}$ is the denominator in Eq.~(\ref{defQ}), $\langle \Delta Q^2 \rangle/Q^2= \langle \Delta B^2 \rangle/B^2+\langle \Delta\Sigma^2 \rangle/\Sigma^2$, and $\langle \Delta B^2 \rangle/B^2 \simeq [3\Delta(k)]^{-1}$ [see Eq.~(\ref{Berror}) below] and $\langle \Delta\Sigma^2 \rangle/\Sigma^2 \simeq 12/N_k^2$ where $N_k$ is the number of k-modes contributing to the estimate of the power spectrum $P(k)$ and $\Delta(k)=4\pi k^3 P(k)$. In other words, the signal to noise of $Q$ is dominated by that of $B$ and the power spectrum can be considered perfectly determined for our purposes.

The bispectrum signal to noise for a given triangle can be written as,
\beq
\left(\frac{S}{N}\right)_{123}\equiv \frac{Q_{123}}{\Delta Q_{123}} \simeq \frac{B_{123}}{\Delta B_{123}},
\label{StoN}
\enq
where the last equality follows from the discussion above. The bispectrum variance in the Gaussian limit can be computed in similar fashion to the standard power spectrum case~\cite{Feldman:1993ky}. For a bispectrum estimator~\cite{Scoccimarro:1997st}

\begin{eqnarray}
\hat{B}_{123} & \equiv & \frac{V_f}{V_{123}}\int_{k_1}\!\!\!\!d^3 q_1\int_{k_2}\!\!\!\!d^3 q_2\int_{k_3}\!\!\!\!d^3 q_3 \;\delta_D(\q_{123}) \nonumber\\
& & \times\ \delta_{\q_1}\delta_{\q_2}\delta_{\q_3} \label{Best}
\end{eqnarray}
where the integration is over the bin defined by $q_i\in(k_i-\delta k/2,k_i+\delta k/2)$, $V_f=(2\pi)^3/V$ is the volume of the fundamental cell, and

\begin{eqnarray}
V_{123} & \equiv & \int_{k_1} \!\!\!\! d^3 q_1\int_{k_2} \!\!\!\! d^3 q_2 \int_{k_3} \!\!\!\! d^3 q_3 \,\delta_D(\q_{123})\nonumber\\
& \simeq & 8\pi^2\ k_1 k_2 k_3\ \delta k^3,
\end{eqnarray}

the variance is~\footnote{Eq.~(\ref{Berror}) corrects Eq.~(A16) in~\cite{Scoccimarro:1997st}.}

\beq
\Delta B^2_{123}=k_f^3\ \frac{s_{123}}{V_{123}}\ P_{tot}(k_1)P_{tot}(k_2)P_{tot}(k_3)
\label{Berror}
\enq
where $s_{123}=6,2,1$ for equilateral, isosceles and general triangles, respectively, and
\beq
P_{tot}(k)  \equiv P(k)+\frac{1}{(2 \pi)^3}\ \frac{1}{ \bar{n}}
\eeq
where the number density $\bar{n}$ accounts for the shot noise.

\begin{figure}
\begin{center}
\includegraphics[width=0.5\textwidth]{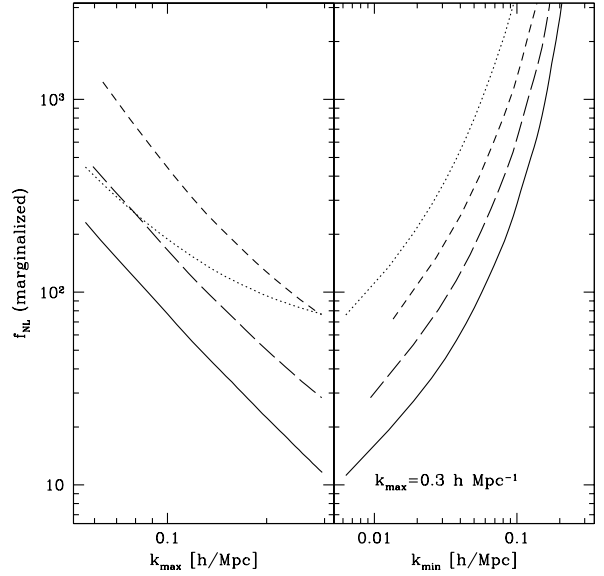}
\caption{Minimum detectable value of $\fnl$ as a function of $k_{\max}$ (left) and $k_{\min}$ (right), after marginalization over bias parameters. The solid line corresponds to an ideal survey with $V=1 \;(\textrm{Gpc}/h)^3$ and no shot noise. The other lines correspond to $V=1 (\Gpc)^3$ with  $\bar{n}=10^{-4} (\kMpc)^3$ (dotted line), $V=0.3 (\Gpc)^3$ with $\bar{n}=3\times 10^{-3} (\kMpc)^3$ (long dashed line) and $V=0.1 (\Gpc)^3$ with $\bar{n}=10^{-3} (\kMpc)^3$ (short dashed line).}
\label{ston_fnl}
\end{center}
\end{figure}

For simplicity, in this section we shall assume that the estimates of the bispectrum are Gaussian distributed  (in practice one can check this assumption for a given survey geometry, see~\cite{Sco00b}), we shall go beyond this in the next section when we obtain bounds on $\fnl$ from the PSCz survey. In the Gaussian approximation, the likelihood for the bispectrum estimates ${\cal L}$ obeys~\cite{MVH97}

\beq
-2\ln {\cal L} = {\rm const}+ \sum_{\rm T} \frac{\left(Q_{\rm obs}-Q_{\rm mod}\right)^2}{\Delta Q_{\rm mod}^2},
\eeq
where T denotes sum over triangles [defined precisely below, Eq.~(\ref{ston})], $Q_{\rm obs}$ is the observed bispectrum and $Q_{\rm mod}$ and $\Delta Q_{\rm mod}^2$ are computed from Eqs.~(\ref{Qgtot}) and (\ref{Berror}) in terms of the model parameters $b_1,\, b_2,\, \fnl$. If observations are consistent with the fiducial model with $b_1=1$, $b_2=0$ and $\fnl=0$, $Q_{\rm obs}=Q^G$ and we have

\beq
-2 \ln {\cal L} =  {\rm const}+ \sum_{i,j=1}^3 \alpha_i\alpha_j\ F_{ij},
\label{GLike}
\eeq
where $\alpha_1=(1-b_1)$, $\alpha_2=b_2/b_1$, $\alpha_3=\fnl$ and the Fisher matrix is given by

\beq\label{ston}
F_{ij} \equiv \sum_{k_1=k_{\min}}^{k_{\max}}\ \sum_{k_2=k_{\min}}^{k_1}\  \sum_{k_3=k_{\min}^*}^{k_2} \ \frac{B^{(i)}_{123}B^{(j)}_{123}}{\Delta B_{123}^2}
\enq
with $k_{\min}^* = \max(k_{\min},|k_1-k_2|)$, and we have assumed that the variance $\Delta B_{123}^2$ is computed only including linear bias.  Here $B^{(1)}_{123}=B^G_{123}$, $B^{(2)}_{123}=\Sigma_{123}$,  and $B^{(3)}_{123}=B^L_{123}/\fnl$, which are respectively the bispectra induced by gravity, non-linear bias and primordial non-Gaussianity. Equation~(\ref{GLike}) is now the standard Gaussian likelihood for the parameters $\alpha_i$ and their error bars  (marginalized over all other $\alpha_j$) are simply given by

\beq
\sigma_i^2=(F^{-1})_{ii}. \label{margerr}
\eeq

\begin{figure}
\begin{center}
\includegraphics[width=0.5\textwidth]{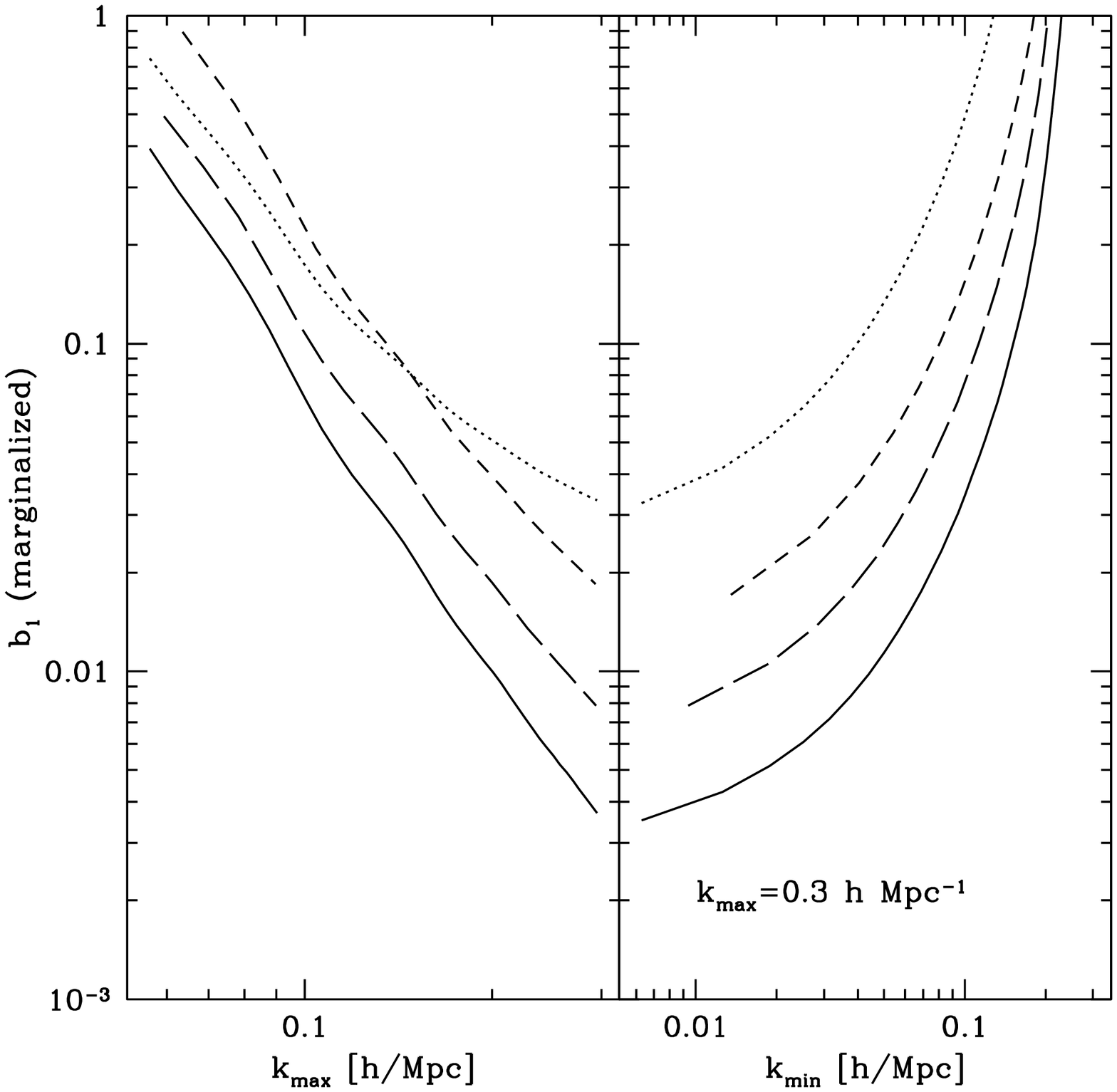}
\includegraphics[width=0.5\textwidth]{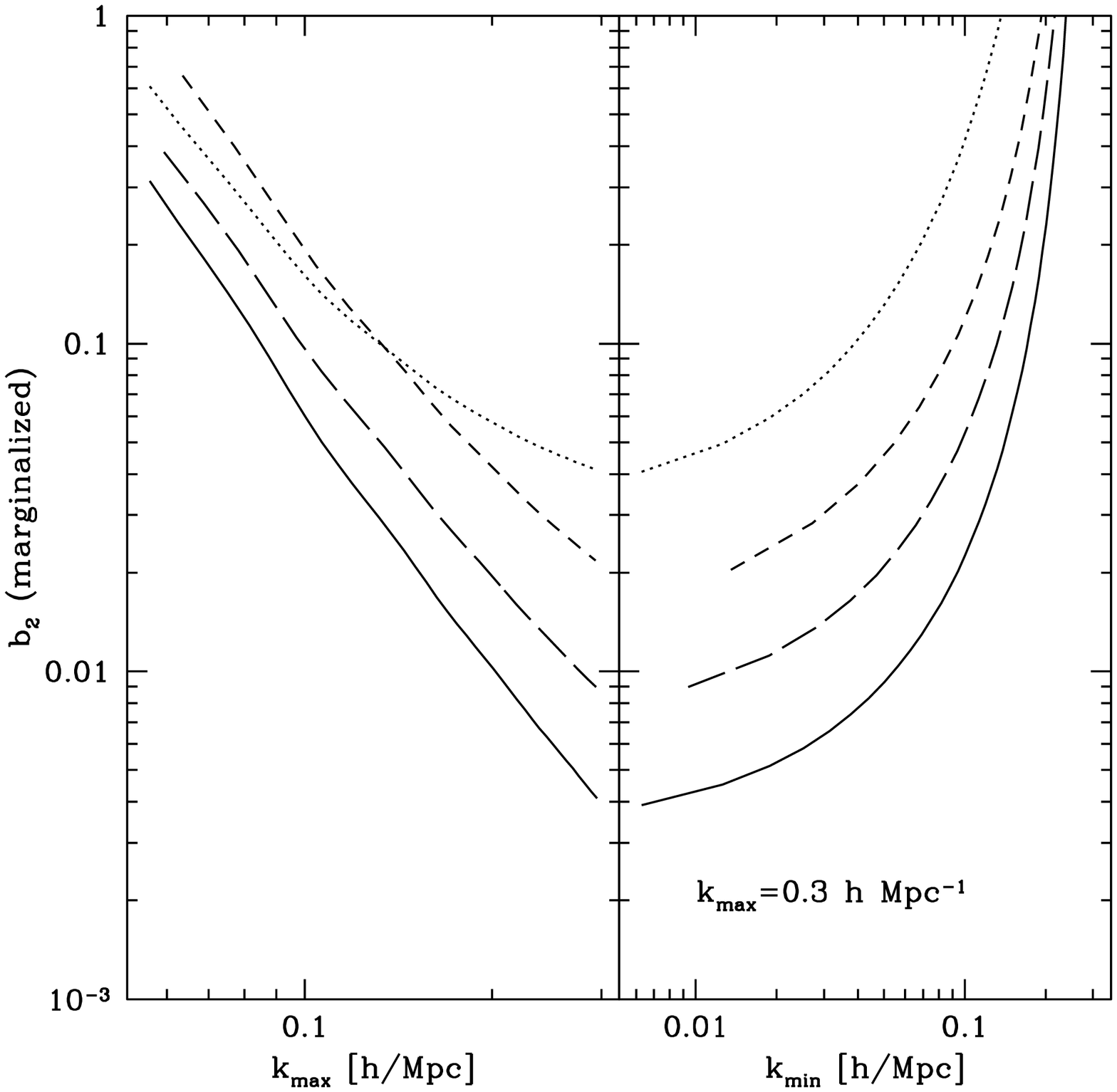}
\caption{Same as Fig.~\ref{ston_fnl} but for the linear ($b_1$) and quadratic bias parameters ($b_2$).}
\label{ston_bias}
\end{center}
\end{figure}

Figure~\ref{ston_fnl} shows the minimal detectable value of $\fnl$, given by $1\sigma$ error bars from Eq.~(\ref{margerr}), for different survey volumes and galaxy number densities. The left panel shows how the $\fnl$ limits improve as we include more triangles towards smaller scales by increasing  $k_{\max}$, whereas the right panel shows the opposite regime, where  $k_{\max}$ is held fixed and one probes larger scales (from right to left) by decreasing $k_{\min}$ all the way up to the fundamental mode of the survey $k_{\min}=k_f$. Figures~\ref{ston_fnl} and~\ref{ston_bias} assume bispectra whose sides are binned with $\delta k= k_f$.

The different lines in Fig.~\ref{ston_fnl} have been chosen to roughly represent the 2dF survey (short-dashed), the main sample (long dashed) and LRG sample (dotted) of the SDSS survey, and a hypothetical survey with the same volume as the LRG sample but with high enough density to make shot noise negligible at $k_{\max}$. In the absence of shot noise (and keeping our ideal survey geometry constant) the minimum detectable value of $\fnl$ scales simply as $V^{-1/2}$.  The scaling with $k_{\max}$  is basically given by the  naive expectation that the constrains on $\fnl$ should be inversely proportional to the square root of the number of modes available, $N_k \propto k_{\max}^3 V$. We thus see from Fig.~\ref{ston_fnl} that an all-sky survey with $\bar{n}\sim 3\times 10^{-3}\, (\kMpc)^3$ up to redshift $z\sim 1$ can probe values of $\fnl$ of order unity. A redshift survey of such a volume may be realistic in the not too distant future~\cite{KAOS}.

Figure~\ref{ston_bias} shows the corresponding results for the bias parameters $b_1$ and $b_2$, showing that 2dF and SDSS surveys should be able to determine the bias parameters to within 1-2 percent accuracy. This is in rough agreement with previous estimates~\cite{MVH97,CSS98}, we will provide a more detailed assessment for the SDSS case in Section~\ref{SDSSF}.

It is worth comparing the right panels of Figs.~\ref{ston_fnl} and~\ref{ston_bias} to see from what scales is most of the signal  coming from. For the case of biasing parameters, the dependence on $k_{\min}$ at large scales is rather weak compared to that for $\fnl$, saying that large-scale triangles contribute significantly more information toward constraining  primordial non-Gaussianity than biasing parameters. This is again a consequence of the {\em scale-dependence} of primordial non-Gaussianity, in fact the signal to noise for an equilateral  triangle of size $k$ is given by ($\Delta\equiv 4\pi k^3 P$)
\beqa
\frac{B^G}{\Delta B} & =  &\frac{4}{7} \sqrt{3\pi\Delta(k)}, \\
\frac{\Sigma}{\Delta B} & =  & \sqrt{3\pi\Delta(k)},
\eeqa
for linear and quadratic bias, respectively, whereas for primordial non-Gaussianity,
\beq
\frac{B^L}{\Delta B}=\sqrt{{48\pi^2 k^3 P_\phi} \over c^2} \ 
 \fnl = 1.4 \times 10^{-4}  \fnl={\rm const},
 \label{StoNfnl}
\eeq
where we have used a scale invariant primordial spectrum normalized to produce the correct level of CMB anisotropies, $(4\pi k^3P_\phi/c^4)^{1/2} \simeq 2.3 \times 10^{-5}$.  Equation~(\ref{StoNfnl}) says that {\em the signal-to-noise per triangle is constant}~\footnote{Here by ``triangle" we mean all triangles with sides  within $k_i\pm k_f/2$, see Eq.~(\ref{Best})}. In other words, for primordial non-Gaussianity given by Eq.~(\ref{fnleq}) the total signal to noise is only decreased at large scales by the decline in the number of triangles, whereas for biasing parameters there is an additional suppression due to the decrease in the {\em rms} fluctuation amplitude at large scales~\cite{FMS93,Scoccimarro:1997st}.

This explains why our constraints on $\fnl$ are about two-orders of magnitude better than those obtained in~\cite{VWHK00}, where the bound on $\fnl$ is derived by translating the constraint on $b_2$ to an effective value of $\fnl$ at $k \sim 0.6~\kMpc$.

\subsection{Including Survey Geometry: SDSS Forecast}
\label{SDSSF}

Let us improve the above treatment, considering a realistic survey geometry with the induced covariance matrix between different triangles. We also include redshift distortions, as calculated by second-order Lagrangian Perturbation Theory (2LPT) from non-Gaussian initial conditions given by Eq.~(\ref{fnleq}), see~\cite{Sco00b} for a comparison of 2LPT against N-body simulations for the redshift-space bispectrum. For biasing, we assume Eq.~(\ref{Qgtot}) still holds in redshift space, which is a reasonable approximation near our fiducial unbiased model. A treatment of bias and primordial non-Gaussianity in the presence of redshift distortions is beyond the scope of this paper.

We consider two survey geometries that approximate the north part of the SDSS survey, a $7,300$ square degree region~\footnote{See {\tt http://www.sdss.org/status} under ``spectroscopy". It corresponds to omitting stripes 17 through 28, and ignoring 76-86 in the south.}  and a second one with $10,400$ square degrees~\footnote{This adds stripes 17 through 28.}.  We don't include the South part of the survey in our analysis, which has a smaller volume and a nearly two-dimensional geometry that complicates the simplified  bispectrum analysis we will do below.  For the radial selection function we use that corresponding to the ``NYU LSS Samples" 10-12~\cite{Blanton03}, and we assume that the angular selection function is unity everywhere inside the survey region, which is a very good approximation.

\begin{table}
\caption{\label{mocks}SDSS mock catalogs (for each geometry) and bispectrum measurements used in the analysis. Cosmological parameters are as in Sect.~\ref{bisplss}, $k_{\min}=0.02 \kMpc$ and $k_{\max}=0.3\kMpc$.}
\begin{ruledtabular}
\begin{tabular}{ccccc}
  $10^{-2}\fnl$ & $N_{\rm mocks}$&  $10^{-3}P_0$  & $N_{\rm triangles}$ & $N_{\rm T}$\\
  0,1,4 & 6080 & 2,5,10  & $7.5\times 10^{10}$ & 1015 \\
\end{tabular}
\end{ruledtabular}
\end{table}

Using a 2LPT code~\cite{Sco00b} with about $42\times 10^6$ particles in a rectangular box of sides $L_i=660,990,~{\rm and}~1320 \Mpc$, we have created about $6\times 10^3$ realizations of each geometry, for Gaussian initial conditions and models with primordial non-Gaussianity with $\fnl=100~{\rm and}~400$ (see Table~\ref{mocks}). In all cases, cosmological parameters are as given in Sect.~\ref{bisplss} and $b_1=1$, $b_2=0$. For each of these realizations, we have measured the redshift-space bispectrum for triangles of all shapes with sides between $k_{\min}=0.02\kMpc$ and $k_{\max}=0.3\kMpc$, giving a total of $N_{\rm triangles}=7.5\times 10^{10}$ triangles. These are binned into $N_{\rm T}=1015$ triangles with a bin size of $\delta k=0.015\kMpc$. The generation of each mock catalog takes about 15 minutes, and has about $4\times 10^5$ galaxies for the smaller area and $5.7 \times 10^5$ galaxies in the larger area case. The redshift-space density field in each mock catalog is then weighed using the FKP procedure~\cite{Feldman:1993ky}, see e.g.~\cite{MVH97,Sco00b} for a discussion in the bispectrum case. We have tried different weights $P_0$ (see Table~\ref{mocks}) to minimize the error bars, the results we present correspond to $P_0=5000~(\Mpc)^3$. The bispectrum in each realization is then measured for all ($\simeq 7.5\times 10^{10}$ binned into 1015) triangles, taking about 2 minutes per realization~\footnote{Timings are for a 1.26 GHz. Pentium III processor.}.

In order to generalize the discussion given above to the case of arbitrary survey geometry, we introduce the bispectrum eigenmodes $\hat{q}_n$~\cite{Sco00b},

\beq
\hat{q}_n = \sum_{m=1}^{N_{\rm T}} \gamma_{mn} \frac{Q_m-\bar{Q}_m}{\Delta Q_m},
\eeq
where $\bar{Q}_m \equiv \langle Q_m \rangle$, $(\Delta Q_m)^2 \equiv \langle (Q_m-\bar{Q}_m)^2 \rangle$. By definition they  diagonalize the bispectrum covariance matrix,
\beq
\langle \hat{q}_n\, \hat{q}_m \rangle = \lambda^2_n \, \delta_{nm},
\eeq
and have signal to noise,

\beq
\left(\frac{S}{N}\right)_n = \frac{1}{\lambda_n}  \left| \sum_{m=1}^{N_{\rm T}} \gamma_{mn} \frac{\bar{Q}_m}{\Delta Q_m}\right| .
\eeq

The eigenmodes are easy to interpret when ordered  in terms of their signal to noise~\cite{Sco00b}.  The best eigenmode (highest signal to noise), say $n=1$, corresponds to all weights $\gamma_{m1}>0$; that is, it represents the overall amplitude of the bispectrum averaged over {\em all} triangles.  The next eigenmode, $n=2$, has $\gamma_{m2}>0 $ for nearly collinear triangles and $\gamma_{m2}<0$ for nearly equilateral triangles, thus it represents the dependence of the bispectrum on triangle shape (see Fig.~\ref{reducedB}).  Higher-order eigenmodes contain further information such as dependence of $Q$ with scale, important to constrain primordial non-Gaussianity~\cite{SFFF01, FFFS01}, see Fig.~\ref{PSCzFig} below for illustration of this point.

If the bispectrum likelihood is Gaussian, we can write down the likelihood as a function of the parameters $\alpha_j$ as,

\beq
{\cal L}(\{\alpha_{j}\}) \propto \prod_{i=1}^{N_{\rm T}}\
P_{i}[\hat{q}_i(\{\alpha_{j}\})],
\label{like}
\eeq
where the $P_{i}(x)$ are all equal and Gaussian with unit variance. We have checked from our Monte Carlo pool that the distribution of $Q$ is indeed Gaussian even at the largest scales we consider. In practice, Gaussianity of $P_{i}$ is not guaranteed at large scales due to the deviations from the central limit theorem by lack of enough uncorrelated triangles~\cite{Sco00b,SFFF01}. If not Gaussian, diagonalization of covariance matrix does not guarantee independence of the eigenmodes, thus ${\cal L}$ does not necessarily factorize as in Eq.~(\ref{like}), but this is a good approximation for small deviations from Gaussianity when the non-Gaussian $P_{i}(x)$ are determined from mock catalogs~\cite{Sco00b}.

\begin{figure}
\begin{center}
\includegraphics[width=0.5\textwidth]{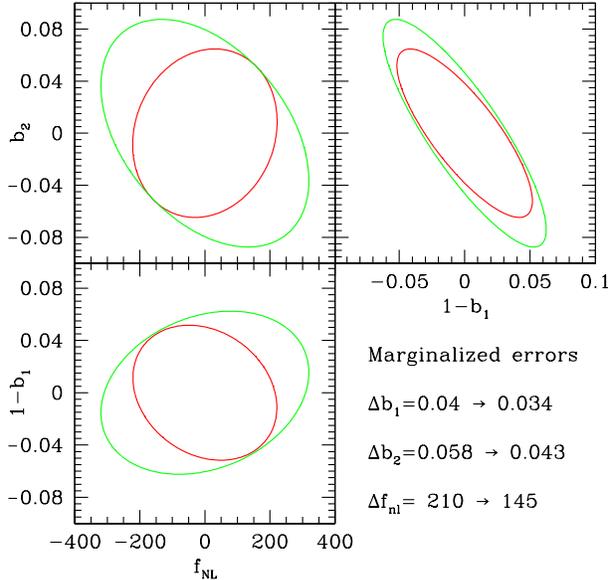}
\caption{Joint 68\% confidence intervals for two parameters at a time, with the third parameter marginalized over. The inner contour corresponds to the larger survey volume case. The lower right panel shows the resulting $1\sigma$ error bars after marginalization, with smaller errors corresponding to the larger volume survey geometry.}
\label{sdsspred}
\end{center}
\end{figure}

We calculate the bispectrum $N_{\rm T} \times N_{\rm T} $ covariance matrix from our $N_{\rm mocks} $ realizations of the survey (see Table~\ref{mocks}) and from that obtain $\gamma_{mn}$ and $\lambda_n$, which gives the ingredients to implement Eq.~(\ref{like}). The results from such likelihood analysis are shown in Fig.~\ref{sdsspred}. Contours denote joint $68\%$ probabilities, two parameters at a time marginalized over the third parameter. The inner contour corresponds to the survey geometry with the larger area. The lower right panel shows the resulting $1\sigma$ error bars after marginalization, smaller uncertainties correspond to the larger volume survey geometry. We have scaled our $\fnl=100$ bispectrum measurements to continuous values of $\fnl$, identical results are obtained by scaling the  $\fnl=400$ mock catalogs.

It is difficult to compare these results to those of the previous section, since they correspond to very  different survey geometries. However, comparing Fig.~\ref{sdsspred} to the long-dashed lines in Figs.~\ref{ston_fnl}-\ref{ston_bias} shows that our more realistic estimates give error bars larger by a factor of 4-5.  There are reasons to expect our ``realistic" treatment to be actually an upper bound to the achievable error bars with a more sophisticated analysis, for the following reasons. First, we only considered the north part of the survey; second we use FKP weighting, which is not optimal at large scales and thus could potentially reduce our sensitivity, particularly to primordial non-Gaussianity; and finally, we have only used closed triangles in Fourier space, due to the lack of translation invariance there is also signal in open configurations.

It is interesting to compare the results of Fig.~\ref{sdsspred} between the two geometries. The larger volume survey leads to an  improvement in marginalized error bars of 20\% for $b_1$, 35\% for $b_2$ and 45\% for $\fnl$. This is more than what one expects for uncorrelated contributions to the constraining power of the survey due to the increased volume ($\sim 20\%$), and is a manifestation in the improvement of the bispectrum covariance matrix due to the narrower survey window function in Fourier space.

\begin{figure}
\begin{center}
\includegraphics[width=0.5\textwidth]{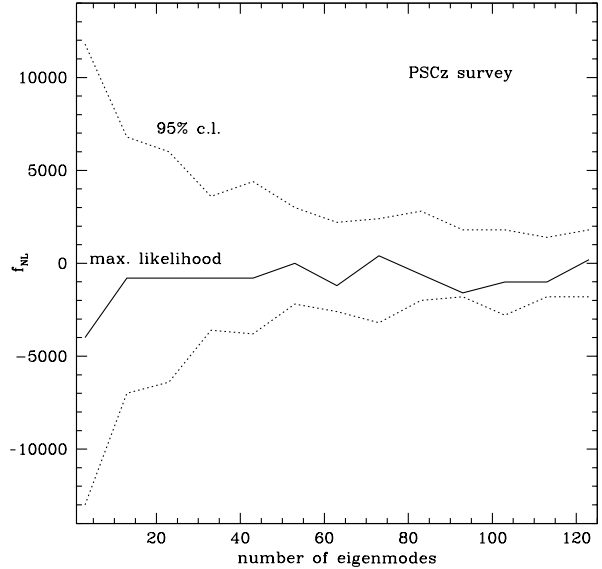}
\caption{$95\%$ confidence limits on $\fnl$ from the PSCz galaxy bispectrum~\cite{FFFS01} after marginalization over bias parameters, as a function of the number of eigenmodes included in the likelihood analysis.}\label{PSCzFig}
\end{center}
\end{figure}

\subsection{$\fnl$ from the PSC$z$ galaxy bispectrum}
\label{pscz}

We now consider constraints on $\fnl$ from the galaxy bispectrum measured in the PSC$z$ survey in~\cite{FFFS01}.  The PSC$z$ survey~\cite{PSCz} is based upon the IRAS Point Source Catalog, the bispectrum measurements we use are based on 13180 galaxies in the range $ 20 \Mpc \le R \le 500 \Mpc $, with galactic latitude $ |b|\ge 10^\circ $, and IRAS 60 micron flux $ f_{60}>0.6 {\rm Jy} $, see~\cite{FFFS01} for more details. We use triangles from  $k_{\min}= 0.05\kMpc$ to $k_{\max}= 0.3\kMpc$, binned into $N_{\rm T}=183$ triangles. We use the eigenmodes and their probability distributions $P_i(x)$ computed from $\sim 10^3$ 2LPT mock catalogs in~\cite{Sco00b}.

Figure~\ref{PSCzFig} shows the $95\%$ confidence limits on $\fnl$ after marginalization over the bias parameters, as a function of the number of eigenmodes (ranked by signal to noise) included in the likelihood analysis. Note how the constraint on $\fnl$ is set by the $n>2$ eigenmodes, which are sensitive to scale dependence of the bispectrum.  The bound on $\fnl$ converges after the best $\sim 100$ eigenmodes are included, since the remaining half of the eigenmodes does not add any significant signal to noise. The $95\%$ confidence limits corresponding  to $123$ eigenmodes are
\beq
-2000 \leq \fnl \leq 1600\ \ \ \ \ (95\%).
\eeq
This is comparable to the constraint from CMB fluctuations before WMAP~\cite{COBEfnl,MAXIMAfnl}, although LSS is sensitive to somewhat smaller scales than the CMB. Our analysis in the previous section suggests that these limits should be improved by about an order of magnitude by the 2dF and SDSS surveys.

\section{Cluster Abundance and Primordial non- Gaussianity}
\label{clus}

The abundance of clusters probes the probability distribution function (PDF) of the matter fluctuations, and it is thus a natural candidate to constrain primordial non-Gaussianity~\cite{LuMa88,KST99,RGS00,Willick00,Matarrese:2000iz,Verde:2000vr}. For large masses, the abundance of clusters depends on the right tail of the PDF which decays exponentially for Gaussian initial conditions. However, before this can be used to place a constrain on $\fnl$, it is necessary to have under control a number of systematic effects.

Even with recent progress in the determination of cosmological parameters, uncertainties  on $\Omega_m$ and in particular $\sigma_8$ can alter the Gaussian abundance prediction enough to make difficult probing the small  levels of primordial non-Gaussianity corresponding to e.g. $\fnl \simeq 100$. In addition, calculation of the mass function from a given PDF is not straightforward, even for Gaussian initial conditions measurements in numerical simulations suffer from systematic  uncertainties of order $10-30\%$ depending on the definition of halo mass~\cite{Jenkins01}. For primordial non-Gaussianity of the type given by Eq.~(\ref{fnleq}), the mass function has been estimated analytically in~\cite{Matarrese:2000iz}, but there has been so far no complementary study using numerical simulations to give a sense of the uncertainties involved.

At a more fundamental level, it is not even clear that one can probe $\fnl$ using rare events such as clusters, given that the tail of the PDF in general {\em is not} determined by the skewness alone, but rather all the higher-order cumulants of the distribution through its generating function~\cite{PTReview}. In other words, the mass function for massive clusters will in general depend on higher-order than quadratic terms in Eq.~(\ref{fnleq}). On the other hand, the first step is to see whether deviations from Gaussianity can be detected at all for models such as Eq.~(\ref{fnleq}), keeping in mind that predictions for the mass function in a fully specified model $\Phi(\phi)$ may be rather different than truncating Eq.~(\ref{fnleq}) to second order in $\phi$.

An important source of uncertainty in comparing theoretical predictions with observations of cluster abundance is the determination of cluster masses. Here we will ignore all the sources of uncertainty mentioned above, and estimate to what accuracy it is necessary to determine cluster masses in order to distinguish a Gaussian from a model with $\fnl=-50,-100$, which being positively skewed gives a larger abundance than the Gaussian case.  For simplicity we assume the Press-Schechter (PS) formula for the mass function~\cite{Press:1973iz,Bond:1990iw},
\beq\label{ps}
\frac{dn\left(M,z\right)}{dM} = \frac{2\bar{\rho}}{M^2} \left| \frac{{\rm d}\ln\sigma_M(z)}{{\rm d}\ln M}\right| \ \nu P(\nu)
\enq
where the mass $M=4\pi\bar{\rho}R^3 /3$ is related to the smoothing scale $R$ and mean density $\bar{\rho}$ by a top hat filter and $\nu=\delta_c / \sigma_M(z)$, with $\delta_c \simeq 1.686$ and $\sigma_M(z)$ the variance of the density field at redshift $z$ smoothed at scale $R$. A more accurate analytic estimate of the mass function in the Gaussian case is given by the Sheth-Tormen (ST) mass function based on ellipsoidal collapse~\cite{Sheth:1999,Sheth:2001,Sheth:2002}, but it is unclear how to generalize it for the non-Gaussian case, therefore we will use Eq.~(\ref{ps}) instead by changing the PDF $P(\nu)$. This has been found to be a reasonable approximation when compared to numerical simulations, at least for large $\nu$ in non-Gaussian models with $\chi^2$ initial conditions~\cite{Robinson:1999se}. We will only deal with {\em ratios} of non-Gaussian to Gaussian abundances, therefore our estimates should be more accurate than differences in the absolute calibration of the mass function. When estimating the total number of clusters expected in the Gaussian case, we will quote both PS and ST values.

The total number of clusters with mass larger than $M$ that can be observed in an all-sky survey between redshifts $z_1$ and $z_2$ is given by,
\beq
N_{cl}(>\!M)=\!\int_{z_1}^{z_2}\!\!\!\!\!\!dz\int \!\!d\Omega\!\int_{M}^{\infty}\!\!\!\!\!\! dM' \frac{dn(M',z)}{dM'}\frac{dV}{dz d\Omega}
\enq
where
\beq
\frac{dV}{dz d\Omega}=\frac{c\, D_c^2(z)}{H_o E(z)}
\enq
with
\beq
D_c=\frac{c}{H_o}\int_0^z\frac{dz'}{E(z')}
\enq
and $E(z)=\sqrt{\Omega_M(1+z)^3+\Omega_\Lambda}$. We assume a  flat Universe with the cosmological parameters given in Sect.~\ref{bisplss}.

\begin{figure}
\begin{center}
\includegraphics[width=0.5\textwidth]{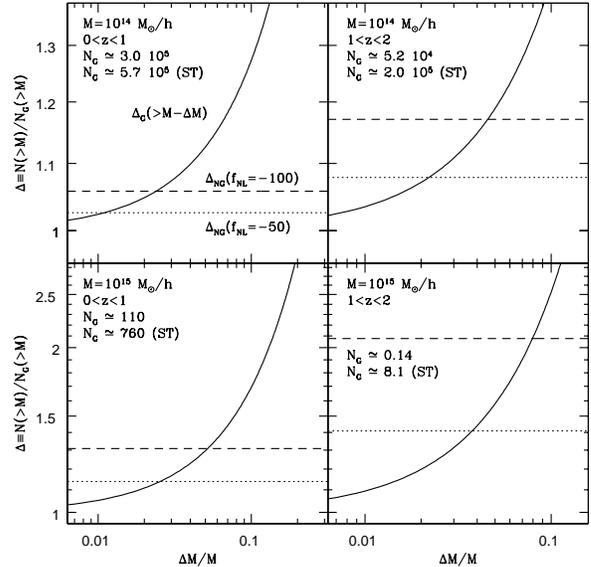}
\caption{The horizontal lines show the excess in the number of clusters in an all-sky survey with redshift limits as shown for masses above $10^{14}$ and $10^{15}\ h^{-1} M_{\odot}$ expected for non-Gaussian primordial fluctuations with $\fnl=-100$ (dashed) and $\fnl=-50$ (dotted). The solid curves show the excess in the number of clusters in the Gaussian case due to an underestimate of the mass limit $M$ by an amount $\Delta M$ as a function $\Delta M/M$. The number of clusters in the Gaussian case $N_G$ for such a survey is given assuming the PS and ST mass functions.}\label{clusters}
\end{center}
\end{figure}

Figure~\ref{clusters} shows the excess in the abundance of clusters due to non-Gaussianity, given by
\beq
\Delta_{NG}(\fnl)\equiv\frac{N^{NG}_{cl}(>M)}{N^{G}_{cl}(>M)}
\enq
for $\fnl=-100$ (dashed) and $\fnl=-50$ (dotted), compared to the excess that may come from an underestimate in the mass determination, given by
\beq
\Delta_{G}(\Delta M/M)\equiv\frac{N^{G}_{cl}(>M-\Delta M)}{N^{G}_{cl}(>M)}
\enq
plotted as a function of $\Delta M/M$. To use Eq.~(\ref{ps}), we have measured the non-Gaussian PDF for the density field from realizations of Eq.~(\ref{fnleq}) on a $512^3$ grid in a box of $1 \Gpc$ a side, smoothing the density field on a scale of $10\Mpc$ (other smoothing lengths in the relevant range do not change the results). Our results in Fig.~\ref{clusters} for the excess abundance are in reasonable agreement with Fig.~8 in~\cite{Matarrese:2000iz}, where an analytic approximation was developed to calculate the mass function instead of using Eq.~(\ref{ps}).

We see from Fig.~\ref{clusters} that in order to probe primordial non-Gaussianity at levels $|\fnl| \la 100$, uncertainties in the mass limit $M$ should be well below $10\%$, assuming perfect knowledge of cosmological parameters, and negligible cosmic variance.  At present, direct mass determinations through weak lensing seems to suffer an absolute uncertainty not below $\Delta M \sim 10^{14} h^{-1} M_\odot$ and larger at high redshift~\cite{Hoekstra:2002cq} with a contribution of distant large scale structure of the order of the $6\%$ alone~\cite{Hoekstra:2001tp}, while the estimates provided by X-ray temperature measurements have {\em statistical} errors of $10-30\%$.

It is interesting to note here that future cluster surveys using the Sunyaev-Zel'dovich effect~\cite{Carlstrom:2002na} require similar accuracy in the determination of the mass threshold for detection. The goal for these observations to be able to probe cosmological parameters is to achieve an accuracy of $5\%$ on the mass limit and $10\%$ on the mass function~\cite{Holder:1999tc,Holder:2001db}. These turn out to be minimal requirements for the use of clusters to probe primordial non-Gaussianity at levels comparable to that within reach of CMB and LSS methods.

\section{Conclusions}

We studied constraints on primordial non-Gaussianity, of the form given by Eq.~(\ref{fnleq}), from measurements of the bispectrum in galaxy redshift surveys.  We find that taking into account the scale-dependence of the bispectrum induced by primordial non-Gaussianity is essential to obtain reliable  constraints on $\fnl$. As a preliminary application, we derived the first constraints on $\fnl$ from LSS using the galaxy bispectrum measured in the PSCz survey, obtaining $-2000 \leq \fnl \leq 1600$ at the 95\% confidence level.

We estimate that the SDSS survey should achieve $68\%$-level constraints for $|\fnl|$ of at least 150-200. The uncertainty in this number is due to the simplifications used in our analysis. Although we have taken into account the geometry in detail by using mock catalogs with the same selection function, there are many features of our analysis that need to be improved for a more accurate assessment. We only consider the north part of the survey, since the south region has a geometry that is nearly two-dimensional and that would invalidate our analysis that assumes the window function of the survey is sufficiently narrow in all directions. To make our analysis more tractable, we have used a weighting scheme that is not necessarily optimal at large scales, for primordial non-Gaussianity this can make a difference since significant information is coming from large scales. For these reasons, our estimate is likely to be an upper limit to the value of $\fnl$ to be probed by SDSS.

In addition, we showed using simple signal to noise estimates that an all-sky survey with $\bar{n}\sim 3\times 10^{-3}\, (\kMpc)^3$ up to redshift $z\sim 1$ can probe values of $\fnl$ of order unity. A redshift survey of such a volume may be realistic in the not too distant future~\cite{KAOS}.

We have also studied the use of cluster abundance to constrain primordial non-Gaussianity. At present, uncertainties appear too large to be able to compete with CMB or LSS, but this can change if cluster masses can be determined with an accuracy of a few percent, and other systematic errors in our theoretical understanding improve.

\acknowledgements
We thank B. Jain, R.K. Sheth, M. Takada and L. van Waerbeke for useful discussions. RS is supported by grants NSF PHY-0101738, and NASA NAG5-12100. Our mock catalogs were created using the NYU Beowulf cluster supported by NSF grant PHY-0116590.

\end{document}